\documentclass[prl,twocolumn,amsmath,showpacs]{revtex4}
\usepackage{graphicx}

\renewcommand\Im{\operatorname{Im}}
\renewcommand{\vec}[1]{\boldsymbol#1}
\def\n{\text{n}}
\def\s{\text{s}}
\def\sd{\text{sd}}

\begin{document}

\title{Quantum mechanical limitations to spin diffusion in the unitary Fermi gas}
\author{Tilman Enss}
\affiliation{Physik Department, Technische Universit\"at M\"unchen,
  D-85747 Garching, Germany} 
\author{Rudolf Haussmann}
\affiliation{Fachbereich Physik, Universit\"at Konstanz,
  D-78457 Konstanz, Germany}

\begin{abstract}
  We compute spin transport in the unitary Fermi gas using the
  strong-coupling Luttinger-Ward theory.  In the quantum degenerate
  regime the spin diffusivity attains a minimum value of $D_\s \simeq
  1.3\,\hbar/m$ approaching the quantum limit of diffusion for a
  particle of mass $m$.  Conversely, the spin drag rate reaches a
  maximum value of $\Gamma_\sd \simeq 1.2\,k_BT_F/\hbar$ in terms of
  the Fermi temperature $T_F$.  The frequency-dependent spin
  conductivity $\sigma_\s(\omega)$ exhibits a broad Drude peak, with
  spectral weight transferred to a universal high-frequency tail
  $\sigma_\s(\omega\to\infty) = \hbar^{1/2}C/3\pi(m\omega)^{3/2}$ proportional to
  the Tan contact density $C$.  For the spin susceptibility
  $\chi_\s(T)$ we find no downturn in the normal phase.
\end{abstract}

\pacs{67.85.Lm, 05.30.Fk, 05.60.Gg, 51.20.+d}
\maketitle

%%%%%%%%%%%%%%%%%%%%%%%%%%%%%%%%%%%%%%%%%%%%%%%%%%%%%%%%%%%%%%%%%%%%%%%%

The excitation and decay of spin currents plays an important role in
many areas of condensed matter physics, including the development of
electronic devices based on spin transport.  While the Coulomb
interaction does not affect electrical currents in a uniform system
\cite{ziman1960}, it transfers momentum between spin-up and down
particles and thereby dampens the spin current.  Understanding the
mechanism of spin drag and spin diffusion quantitatively is important
for an effective control of spin currents; however, in solids this is
often complicated by the presence of impurities and lattice effects.
Ultracold atomic Fermi gases provide an extremely clean experimental
realization to study the effect of the two-particle interaction alone
\cite{bloch2008}.  If the interactions are short-ranged and the
scattering length is much larger than the particle spacing the
results are universal and apply to a wide range of models, including
dilute nuclear matter.  

The spin diffusivity $D_\s$ measures how quickly a spin current levels
out a gradient in the spin density.  In a strongly interacting Fermi
gas $D_\s$ decreases as the temperature is lowered into the quantum
degenerate regime and reaches a minimum near the Fermi temperature
$T_F$, before increasing again at even lower temperatures in the
superfluid phase.  The minimum value of $D_\s$ in the strong-coupling
region can be understood qualitatively as a consequence of the
uncertainty principle: the mean-free path in a gas cannot become
shorter than the mean particle spacing in the absence of localization
\cite{sachdev1999}, which translates into a quantum bound $D_\s
\gtrsim \hbar/m$ for particles of mass $m$.  For a strongly
interacting Fermi gas of trapped $^6$Li atoms a spin diffusivity $D_\s
\geq 6.3(3)\,\hbar/m$ has recently been measured
\cite{sommer2011etal}.  Very low spin diffusion is found also in
graphene \cite{mueller2011}, while spin Coulomb drag in GaAs quantum
wells yields a value of $D_\s \gtrsim 500\,\hbar/m$ \cite{weber2005}.

\begin{figure}[b]
  \centering
  \includegraphics[width=\linewidth]{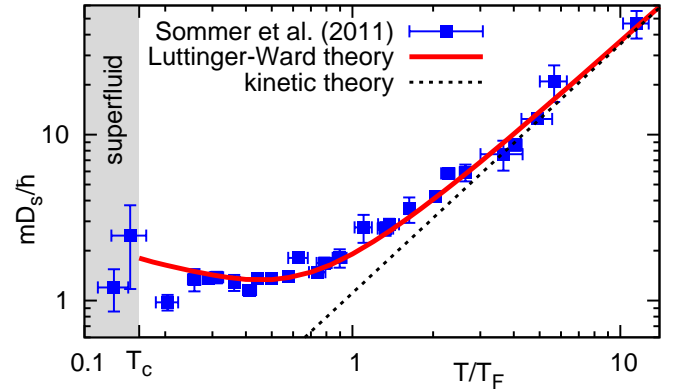}
  \caption{(Color online) Spin diffusivity $D_\s$ vs.\ reduced
      temperature $T/T_F$ (solid red line) in the normal phase, $T>T_c
      \simeq 0.16\,T_F$.  The experimental data \cite{sommer2011etal}
      (blue squares) for the trapped gas are rescaled down by a factor
      of $4.7$ to compensate for the effect of the trapping potential.
      The dashed black line is the result from kinetic theory, $D_\s =
      1.1\,(T/T_F)^{3/2} \hbar/m$.}
  \label{fig:diff}
\end{figure}

The determination of $D_\s$ near its minimum in the strongly
interacting regime, and more generally the question of whether quantum
mechanics imposes universal lower bounds on the transport
coefficients, is a key challenge in many-body physics.  Recent
progress comes from the anti-de-Sitter and conformal field theory
correspondence which maps a strongly coupled field theory to an
equivalent weakly coupled gravitational theory, where calculations are
feasible.  It gives a lower quantum bound for the internal friction of
mass flow, expressed as the ratio of shear viscosity to entropy
$\eta/s\geq \hbar/4\pi k_B$, in certain relativistic field theories
\cite{policastro2001etal}.  Quantum limited friction, or perfect
fluidity \cite{schaefer2009etal}, has been found to be almost
satisfied in very different physical situations ranging from
quark-gluon plasmas to ultracold atomic gases \cite{enss2011,
  cao2011etal, enss2012critical}.  It remains an open question whether
a similar bound exists for spin diffusion in nonrelativistic systems
\cite{son2008ads}.  

In this work we present a strong-coupling calculation of the spin
diffusivity $D_\s$ in the unitary Fermi gas.  At infinite scattering
length it saturates the unitarity bound on the scattering cross
section and is one of the most strongly interacting systems known; it
is also the only known example of a nonrelativistic interacting
scale-invariant fluid.  The unitary gas becomes superfluid below the
transition temperature $T_c \simeq 0.16\,T_F$ \cite{ku2012}; here we
focus on the normal phase above $T_c$ where transport experiments are
available and where the most interesting features occur.  In
Fig.~\ref{fig:diff} our result for the diffusivity is shown by the
solid red line, and we find a minimum value of $D_\s \simeq
1.3\,\hbar/m$ at a temperature of about $T=0.5\,T_F$.  To our
knowledge this is the lowest value achieved to date for a strong
two-particle interaction.  Recent experimental data for the trapped
unitary gas \cite{sommer2011etal} are shown as the blue squares (see
caption), and we obtain remarkable agreement for all temperatures in
the normal phase.

In the high-temperature limit where $D_\s \gg \hbar/m$ is much larger
than the quantum limit our calculation agrees with the predictions of
Boltzmann kinetic theory \cite{duine2010, sommer2011etal, bruun2011,
  bruun2011spin} (dashed black line).  In the strongly interacting
region near $T_c$, however, the fermions cease to be well-defined
quasiparticles \cite{haussmann2009, gaebler2010} and the Boltzmann theory
is not applicable.  Therefore, we employ the strong coupling
Luttinger-Ward theory to compute spin transport.  The Luttinger-Ward
(or 2PI) formalism \cite{luttinger1960, baym1961} is based on the
self-consistent $T$ matrix for repeated particle-particle scattering
and becomes exact at high temperatures.  In the most interesting
regime near $T_c$ and unitarity there is no small parameter to
estimate its accuracy.  Instead, a comparison with experiment shows
that it accurately describes both the normal and the superfluid phase
of the BEC-BCS crossover problem \cite{haussmann1993etal}: the values
for $T_c/T_F = 0.16(1)$ and the Bertsch parameter $\xi=0.36(1)$ agree
within error bounds with precision experimental \cite{ku2012} and
diagrammatic Monte Carlo \cite{vanhoucke2012} results.  We have
devised a framework which includes all diagrams needed to exactly
fulfill the conservation laws including scale invariance
\cite{enss2011} and the Tan relations \cite{enss2012critical}.

The Luttinger-Ward
theory has recently been extended to compute transport coefficients in
linear response using the Kubo formula: this gives access to the
frequency-dependent shear viscosity of the unitary Fermi gas, which
was found to satisfy the exact viscosity sum rule \cite{taylor2010,
  enss2011}.  We now extend this work to the case of spin transport in
order to explain the recent experiment by Sommer \emph{et al.}\
\cite{sommer2011etal}, and we proceed as follows: first we compute the
frequency-dependent spin conductivity $\sigma_\s(\omega)$ of the
unitary Fermi gas.  The dc value $\sigma_\s = \sigma_\s(\omega=0)$
determines the spin drag rate $\Gamma_\sd = n/m\sigma_\s$ at density
$n$, which is the rate of momentum transfer between atoms of opposite
spin.  We then compute the spin susceptibility $\chi_\s
= \partial(n_\uparrow - n_\downarrow) / \partial(\mu_\uparrow -
\mu_\downarrow)$ which characterizes the magnetic properties of the
system \cite{stringari2009, duine2010}.  Finally, we determine the
spin diffusivity shown in Fig.~\ref{fig:diff} by the Einstein relation
$D_\s=\sigma_\s / \chi_\s$.

%%%%%%%%%%%%%%%%%%%%%%%%%%%%%%%%%%%%%%%%%%%%%%%%%%%%%%%%%%%%%%%%%%%%%%%%

The strongly interacting two-component Fermi gas is described by the
grand canonical Hamiltonian
\begin{align*}
  \mathcal H
  = \sum_{\vec k,\sigma} (\varepsilon_{\vec k}-\mu_\sigma)
  c_{\vec k\sigma}^\dagger c_{\vec k\sigma}
  + \frac{g_0}{V} \sum_{\vec k,\vec k',\vec q}
  c_{\vec k\uparrow}^\dagger c_{\vec k'\downarrow}^\dagger
  c_{\vec k'-\vec q\downarrow} c_{\vec k+\vec q\uparrow}
\end{align*}
where $\varepsilon_{\vec k}=\vec k^2/2m$ ($\hbar\equiv 1$) is the free
particle dispersion and $\mu_\sigma$ the chemical potential for the
$\sigma=\;\uparrow,\downarrow$ components.  The $s$-wave contact
interaction $g_0$ acts only between different fermion species at low
temperatures.  The bare interaction is singular in the ultraviolet
\cite{bloch2008} and needs to be regularized; the renormalized
coupling $g=4\pi\hbar^2a/m$ determines the $s$-wave scattering length
$a$.

The transport coefficients are obtained from the microscopic model via
the retarded number-current/spin-current correlation function
\begin{multline}
  \label{eq:spincorr}
  \chi_\text{jn/js}(\vec q,\omega) = \frac{i}{\hbar} \int_0^\infty dt
  \int d^3x \, e^{i(\omega t-\vec q\cdot \vec x)}\\
  \times \left\langle\Bigl[(j_\uparrow^z \pm j_\downarrow^z)(\vec x,t),
    (j_\uparrow^z \pm j_\downarrow^z)(\vec 0,0)\Bigr]\right\rangle .
\end{multline}
The spin selective current operators in Fourier representation are
given by $\vec j_\sigma(\vec q) = V^{-1} \sum_{\vec k} (\hbar\vec k/m) c_{\vec
  k-\vec q/2,\sigma}^\dagger c_{\vec k+\vec q/2,\sigma}$.  The
correlation function determines the conductivity
\begin{align}
  \label{eq:spincond}
  \sigma_\text{n/s}(\omega) = \lim_{q\to0} 
  \frac{\Im\chi_\text{jn/js}(\vec q,\omega)} {\omega}
\end{align}
which measures the relaxation of a global number/spin current at
frequency $\omega$.  The total response integrated over all
frequencies is proportional to the particle density by the number/spin
$f$-sum rule \cite{abrikosov1975, enss2012sumrule}
\begin{align}
  \label{eq:fsum}
  \int_{-\infty}^\infty \frac{d\omega}{\pi}\,
  \sigma_\text{n/s}(\omega) = \frac nm.
\end{align}
For a momentum-conserving interaction the particle
current cannot decay and $\sigma_\n(\omega) = \pi n \delta(\omega)/m$.
In contrast, scattering transfers momentum between $\uparrow$ and
$\downarrow$ particles so that the spin current relaxes and
$\sigma_\s(\omega)$ has a nontrivial structure.

We compute the current correlation function \eqref{eq:spincorr} using
field theoretical methods and Feynman diagrams in the Matsubara
formalism \cite{abrikosov1975}.  The current operator $j^z =
j_\uparrow^z\pm j_\downarrow^z$ implies a current response vertex
$J_{\sigma\sigma'} = J_{\sigma\sigma'}^0 + J_{\sigma\sigma'}^\text{MT}
+ J_{\sigma\sigma'}^\text{AL}$ in the Feynman diagrams which splits
into three contributions \cite{baym1961, enss2011} ($\sigma,\sigma'$
are the spin indices of incoming and outgoing fermion lines).  The
first term is the bare number (spin) current vertex
$J_{\sigma\sigma'\n}^0(\vec p) = p_z \tau_{\sigma\sigma'}^0$
($J_{\sigma\sigma'\s}^0(\vec p) = p_z \tau_{\sigma\sigma'}^3$) with
the $\ell=1$ partial wave component of the momentum $\vec p$ and Pauli
matrices $\tau^j$.  The other two terms are current vertex corrections
which are required to fulfill the conservation laws.  The
Maki-Thompson (MT) contribution describes direct scattering between
quasiparticles while the Aslamazov-Larkin (AL) term captures the
induced current of fermion pairs, or molecules (for details see
Ref.~\cite{enss2011}).  For a mass current both $\uparrow$ and
$\downarrow$ fermions move in the same direction and induce a current
of pairs, leading to a sizeable AL term.  In contrast, for a spin
current $\uparrow$ and $\downarrow$ atoms move in opposite directions
\cite{sommer2011etal} and no pair current is induced.  Hence, the
Aslamazov-Larkin correction to the spin current vanishes exactly in
the spin balanced case, $J_{\sigma\sigma'\s}^\text{AL} = 0$, which
constitutes an important simplification.

We solve the self-consistent equation for the fully dressed current
vertex $J_{\sigma\sigma'}$ by iteration and obtain the current
correlation function \eqref{eq:spincorr} via the Kubo formula
\cite{enss2011}.  Since the correlation function
$\chi_\text{jn/js}(q=0,i\omega_m)$ is evaluated at discrete imaginary
Matsubara frequencies $i\omega_m$, we must perform an analytic
continuation in order to obtain the physically relevant correlation
function $\chi_\text{jn/js}(\omega)$ for real frequencies $\omega$.
We use Pad\'e approximants and find that the continuation is robust at
low temperatures if we vary the number of Matsubara frequencies, and
it yields the correct high-frequency tail (see below).  Specifically,
we oversample the Matsubara data twice with a spline fit and use the
first five Matsubara frequencies in order to extract the spin drag
rate $\Gamma_\sd$.  We validate our strong coupling calculation by
confirming that $\sigma_\s(\omega)$ indeed fulfills the spin $f$-sum
rule \eqref{eq:fsum} within $1\%$.  Since we have constructed the
formalism to satisfy the sum rules exactly, this quantifies the
numerical accuracy of our self-consistent solution and the analytical
continuation.

%%%%%%%%%%%%%%%%%%%%%%%%%%%%%%%%%%%%%%%%%%%%%%%%%%%%%%%%%%%%%%%%%%%%%%%%

\begin{figure}[t]
  \centering
  \includegraphics[width=\linewidth]{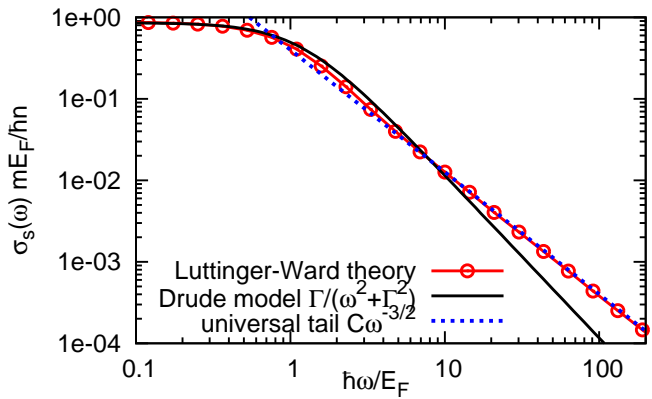}
  \caption{(Color online) Spin conductivity $\sigma_\s(\omega)$ (in
    units of $\hbar n/mE_F$) vs.\ frequency (red circles) at
    $T=0.5\,T_F$.  The Drude model (solid black line) has the same
    total spectral weight as $\sigma_\s(\omega)$ given by the spin
    $f$-sum rule.  Part of the spectral weight is transferred from
    lower frequencies into a universal high-frequency tail (dotted
    blue line) $\sigma_\s(\omega\to\infty) = \hbar^{1/2}C/3\pi(m\omega)^{3/2}$
    with Tan contact density $C=0.086\,k_F^4$ \cite{enss2011}.}
  \label{fig:dyn}
\end{figure}

\textit{Spin conductivity.}---The resulting spin conductivity
$\sigma_\s(\omega)$ is shown in Fig.~\ref{fig:dyn} for reduced
temperature $T/T_F = 0.5$ where it has the lowest dc value $\sigma_\s
= 0.8\,n/m$ (red circles).  In a Drude model the conductivity would
assume a form $\sigma_\s^\text{Drude}(\omega) = (n/m) \Gamma_\sd /
(\omega^2+\Gamma_\sd^2)$ (solid black line) with total spectral weight
given by the sum rule.  The spin drag rate $\Gamma_\sd$ is a parameter
which we determine from the dc limit $\sigma_\s = n/m\Gamma_\sd$ of
our full numerical solution.  We find that the true
$\sigma_\s(\omega)$ deviates from the Drude model for $\omega \gtrsim
E_F$: spectral weight is transferred from the region $\omega \lesssim
8\,E_F$ to higher frequencies where it forms a power-law tail
$\sigma_\s(\omega\to\infty) \sim \omega^{-3/2}$ (dotted blue line in
Fig.~\ref{fig:dyn}).

The high-frequency response generally depends on the nonuniversal
short-distance behavior of the interatomic potential.  However, for a
broad Feshbach resonance as in $^6$Li \cite{bloch2008} this potential
has a range much shorter than the particle spacing, $k_F|r_e| \ll 1$,
and becomes effectively a contact interaction.  In this case the
correlation functions exhibit universal power-law tails in the
high-frequency range $\max(E_F,k_BT)/\hbar \ll \omega \ll \hbar/(mr_e^2)$
\cite{braaten2012} which depend only on the Tan contact density $C$
\cite{tan2008large}.  In the high-frequency limit the exact transport
equations can be solved analytically in a manner analogous to the
viscosity response \cite{enss2011}, and we obtain the universal spin
conductivity tail
\begin{align}
  \label{eq:tail}
  \sigma_\s(\omega\to\infty) = \frac{\hbar^{1/2}C}{3\pi(m\omega)^{3/2}}
\end{align}
in agreement with the result from the operator product expansion
\cite{hofmann2011}.  Similar tails appear in other transport
properties such as the viscosity \cite{taylor2010, enss2011,
  hofmann2011, goldberger2012etal}.  The value for the Tan contact
density $C = 0.0863\,k_F^4$ at $T/T_F = 0.5$ extracted from the tail
of $\sigma_\s(\omega)$ agrees better than $1\%$ with the value $C =
0.0860\,k_F^4$ from the tail of the momentum distribution $n_k \sim
Ck^{-4}$ \cite{enss2011}.  A similar behavior of $\sigma_\s(\omega)$
is observed for all temperatures $T\geq T_c$.

\begin{figure}[t]
  \centering
  \includegraphics[width=\linewidth]{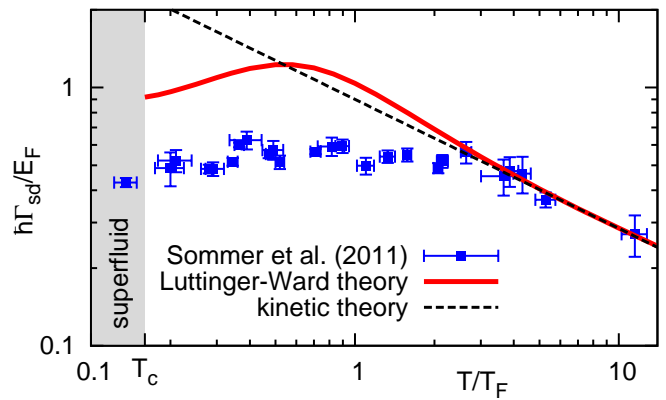}
  \caption{(Color online) Spin drag rate $\Gamma_\sd$ (in units of
    $E_F/\hbar$) vs.\ reduced temperature $T/T_F$ (solid red line).
    The experimental data \cite{sommer2011etal} (blue squares) for a
    trapped gas are rescaled up by a factor of $5.3$ to compensate for
    the effect of the trapping potential.  The dashed black line is
    the result from kinetic theory, $\Gamma_\sd =
    0.9\,(T/T_F)^{-1/2}E_F/\hbar$.}
  \label{fig:drag}
\end{figure}

We now turn to the dc limit and plot the spin drag rate
$\Gamma_\sd=n/m\sigma_\s$ in Fig.~\ref{fig:drag} (solid red line).
The spin drag has a maximum value of $\Gamma_\sd \approx
1.2\,E_F/\hbar$ in the quantum degenerate regime around $T/T_F = 0.5$
and decreases both for lower and higher temperatures.  In the
high-temperature limit of a classical gas the Luttinger-Ward transport
equations can be solved analytically to leading order in the fugacity
\cite{enss2011}, and we obtain $\Gamma_\sd = (32\sqrt 2/9\pi^{3/2})
(T/T_F)^{-1/2} E_F/\hbar = 0.9\, (T/T_F)^{-1/2} E_F/\hbar$ for $T\gg
T_F$ in agreement with Boltzmann kinetic theory \cite{sommer2011etal,
  bruun2011}.  The fact that the numerical solution at large
temperatures agrees with the analytical result for $T\gg T_F$ is a
nontrivial validation of our analytical continuation procedure.

The measured spin drag rate in a trapped unitary Fermi gas
\cite{sommer2011etal} (blue squares in Fig.~\ref{fig:drag}) has the
same qualitative behavior as our numerical data, with a broad maximum
between $T/T_F=0.4\ldots 0.8$.  Note that the absolute spin drag rate
cannot be directly compared to our calculation for the uniform system:
the solution of the transport equation depends on the trap geometry
and the velocity profiles of $\uparrow$ and $\downarrow$ particles in
the trap \cite{sommer2011etal, bruun2011spin, goulko2011etal}.  For a
quadratic velocity profile in a harmonic trap the spin drag rate
$\Gamma_\sd^\text{trap} = \Gamma_\sd/\alpha$ is rescaled by a constant
factor $\alpha=2^{5/2}$ in the high-temperature limit (see
supplementary information of Ref.~\cite{sommer2011etal}).  In the
experiment a factor of $\alpha=5.6(4)$ is found, and we obtain the
best fit at high temperatures for $\alpha=5.3$.  In the quantum
degenerate regime $T \lesssim T_F$ the assumption of a uniform
quadratic velocity profile breaks down: in the center of the trap a
large spin drag leads to slow spin motion, while the spins in the
weakly interacting wings move rapidly.  The velocity profile thus
becomes nonuniform and $\alpha$ acquires a temperature dependence.
In Fig.~\ref{fig:drag} the calculation for the uniform system and the
rescaled trap-averaged data differ for $T\lesssim T_F$, and the
scaling factor starts to deviate from the high-$T$ estimate
$\alpha=5.3$.

%%%%%%%%%%%%%%%%%%%%%%%%%%%%%%%%%%%%%%%%%%%%%%%%%%%%%%%%%%%%%%%%%%%%%%%%

\textit{Spin susceptibility.}---We shall compute and discuss the spin
susceptibility $\chi_\s$ in order to find $D_\s=\sigma_\s/\chi_\s$.
Both the spin susceptibility $\chi_\s = \partial(n_\uparrow -
n_\downarrow) / \partial(\mu_\uparrow - \mu_\downarrow)$ and the
normalized compressibility $\chi_\n = n^2\kappa = \partial(n_\uparrow
+ n_\downarrow) / \partial(\mu_\uparrow + \mu_\downarrow)$ are
obtained from the number/spin correlation function
\begin{align*}
  \chi_\text{n/s} = \frac{i}{\hbar} \int_0^\infty dt\,d^3x
  \left\langle\Bigl[(n_\uparrow \pm n_\downarrow)(\vec x,t), (n_\uparrow \pm
    n_\downarrow)(\vec 0,0)\Bigr]\right\rangle .
\end{align*}
The spin-selective particle number operator in Fourier representation
reads $n_\sigma(\vec q) = V^{-1} \sum_{\vec k} c_{\vec k-\vec
  q/2,\sigma}^\dagger c_{\vec k+\vec q/2,\sigma}$.  In the
Luttinger-Ward formulation the bare number (spin) density vertex has
the form $N_{\sigma\sigma'\n}^0=\tau_{\sigma\sigma'}^0$
($N_{\sigma\sigma'\s}^0=\tau_{\sigma\sigma'}^3$).  For the dressed
number density vertex $N_\n$ both MT and AL vertex corrections
contribute, while the AL term again vanishes for $N_\s$ in the spin
balanced case.
\begin{figure}[t]
  \centering
  \includegraphics[width=\linewidth]{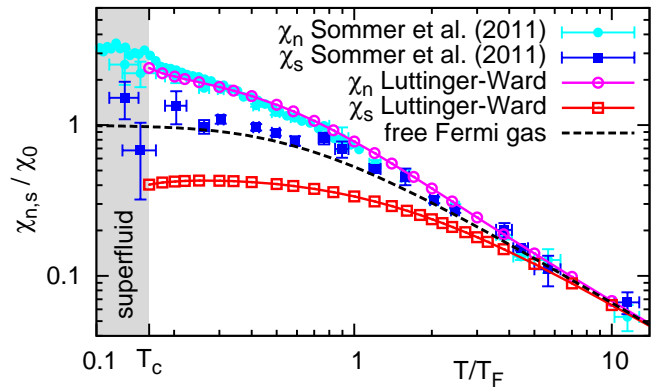}
  \caption{(Color online) Compressibility $\chi_\n$ (circles) and spin
    susceptibility $\chi_\s$ (squares) vs.\ reduced temperature
    $T/T_F$.  The experimental data \cite{sommer2011etal} (full
    symbols) are compared to our Luttinger-Ward calculation
    (open symbols).  The dashed black line is the susceptibility of
    the free Fermi gas.}
  \label{fig:susc}
\end{figure}
In order to obtain the static susceptibility $\chi_\text{n/s}$ we
calculate the susceptibility $\chi_\text{n/s}(i\omega_m=0)$ for zero
Matsubara frequency; note that an analytical continuation is not
needed here.  The static limit of the related \emph{current}
correlation function $\chi_\text{jn/js} =
\chi_\text{jn/js}(i\omega_m=0)=n/m$ is fixed by the exact $f$-sum rule
\eqref{eq:fsum}, and our numerical computation fulfills this sum rule
within $1\%$ (see above).  We therefore expect our results for
the static susceptibilities $\chi_\text{n/s}$ to be of the same
accuracy.

The susceptibility of the free Fermi gas is $\chi_{\n,\s}^0 = n/k_BT$ for
$T\gg T_F$ (Curie-Weiss) and $\chi_{\n,\s}^0 = \chi_0 = 3n/2E_F$ for
$T\to0$ in the Fermi liquid phase (dashed black line in
Fig.~\ref{fig:susc}).  In the unitary Fermi gas the attractive
interaction leads to a compressiblity $\chi_\n$ twice as large as
$\chi_\n^0$ in the quantum degenerate regime near $T_c$ (open magenta
circles), in very good agreement with the experimental data
\cite{sommer2011etal} (full cyan circles) and with a
non-self-consistent diagrammatic approach \cite{palestini2012}.
Conversely, we find that the spin susceptibility $\chi_\s$ remains
below $\chi_\s^0$ and exhibits a maximum of about $\chi_\s \simeq
  0.4\,\chi_0$ at $T/T_F=0.3$ (open red squares).  The spin
susceptibility is expected to vanish as $\exp(-2\Delta/k_BT)$ deep in the
superfluid phase with gap $\Delta$, where an infinitesimal magnetic
field gradient cannot break pairs.  The proposed pseudogap scenario
\cite{gaebler2010, palestini2012} predicts a pronounced drop of
$\chi_\s$ at a pair-breaking scale $T^* > T_c$.  Our data, which fully
include the attractive branch, remain nearly constant down to $T/T_F
\simeq 0.2$.  This indicates that the scales $T^*$ and $T_c$ are very
close in the unitary Fermi gas.  The measured $\chi_\s$ (full blue
squares in Fig.~\ref{fig:susc}) also shows no downturn and can be
described in a Fermi liquid picture despite the large value for
$T_c/T_F \simeq 0.16$, in agreement with a recent quantum Monte Carlo
and experimental study \cite{nascimbene2011}.  Note that a finite
condensate fraction can lead to significantly lower values for
$\chi_\s$ \cite{sanner2011speckle}.

In the experiment by Sommer \emph{et al.}\ \cite{sommer2011etal}
$\chi_\s$ is determined from a combination of the local spin density
gradient and the trap-averaged center of mass motion.  Hence, the
measured $\chi_\s$ deviates at low temperatures from the calculation
of the uniform system (cf.\ Fig.~\ref{fig:susc}) for similar reasons
as discussed above for $\Gamma_\sd$.  Remarkably, we find that the
differences in $\Gamma_\sd$ and $\chi_\s$ cancel and lead to a very
good agreement in the spin diffusivity $D_\s = n/m\Gamma_\sd \chi_\s$
shown in Fig.~\ref{fig:diff} above.

In conclusion, our strong coupling calculation of spin transport
explains the behavior of the spin diffusivity $D_\s$ seen in
experiment \cite{sommer2011etal}, and we find that the diffusivity of
the unitary Fermi gas reaches the quantum limit $\hbar/m$.  This
provides an important constraint for any future spin transport bound
from gravity duals \cite{son2008ads}.  We predict a universal
high-frequency tail of the spin conductivity $\sigma_\s(\omega)$ which
should be accessible experimentally using Bragg spectroscopy for the
dynamic structure factor \cite{hoinka2012}.  It would be desirable to
have local measurements of transport properties in a way similar to
local precision measurements of the thermodynamic properties
\cite{ku2012} and the momentum distribution \cite{drake2012}.

\begin{acknowledgments}
  We thank Johannes Hofmann, Mark Ku, Sergej Moroz, Ariel Sommer,
  Wilhelm Zwerger, and Martin Zwierlein for fruitful discussions.
\end{acknowledgments}

\end{document}